\documentclass{PoS}
      \usepackage{bm}

\title{Subleading processes in production of $W^+ W^-$ pairs in proton-proton collisions}

\ShortTitle{Subleading processes in production of $W^+ W^-$}

\author{\speaker{Antoni Szczurek}%
        \thanks{This study was partially supported by the Polish
National Science Centre grant DEC-2013/09/D/ST2/03724.}\\
       Institute of Nuclear Physics PAN, PL-31-342 Cracow, Poland\\
       E-mail: \email{antoni.szczurek@ifj.edu.pl}}

\author{Marta Luszczak\\%
       University of Rzesz\'ow, PL-35-959 Rzesz\'ow, Poland\\
       E-mail: \email{luszczak@ur.edu.pl}}

\abstract{We discuss new subleading processes for inclusive production of 
$W^+ W^-$ pairs. 
We focus mainly on photon-photon induced processes.
We include elastic-elastic, elastic-inelastic, inelastic-elastic 
and inelastic-inelastic contributions. 
The inelastic photon distributions in the proton are 
calculated in two different ways: naive approach used already in 
the literature and using photon distributions by solving special 
evolution equations with the photon being a parton in the proton. 
The results strongly depend on the approach used. 
The resolved photon contribution was calculated in addition
and found to be small.
We also calculate the cross section for single-diffractive production of
$W^+ W^-$ pairs including pomeron and reggeon exchanges
in the Ingelman-Schlein model.
Finally we only mention here about double parton contribution
which is interesting by itself.
}

\FullConference{The European Physical Society Conference on High Energy Physics\\
          22-29 July 2015\\
          Vienna, Austria}

\begin{document}

\section{Introduction}

We shortly review several subleading processes in the production of
$W^+ W^-$ pairs in proton-proton collisions \cite{LSR2015}.
The $\gamma \gamma \to W^+ W^-$ process, one of them, is interesting 
by itself as it can be used to test the Standard Model and any 
other theories beyond the Standard Model.
The exclusive diffractive mechanism of central exclusive production
of $W^+W^-$ pairs in proton-proton collisions at the LHC 
(in which diagrams with an intermediate virtual Higgs boson as well 
as quark box diagrams are included) was discussed 
in Ref.~\cite{LS2012} and turned out to be negligibly small.
The diffractive production and decay of the Higgs boson into 
the $W^+W^-$ pair was also discussed in Ref.~\cite{WWKhoze}. 
The $W^+W^-$ pair production signal would
be particularly sensitive to New Physics contributions in 
the $\gamma \gamma \to W^+ W^-$ subprocess \cite{royon,piotrzkowski}. 
A similar analysis has been considered recently
for $\gamma \gamma \to Z Z$ \cite{Gupta:2011be}. 
Corresponding measurements would be possible at ATLAS or CMS
provided very forward proton detectors are installed. 
We concentrate on inclusive production of $W^+ W^-$ pairs.
The inclusive production of $W^+ W^-$ has been measured recently with 
the CMS and ATLAS detectors \cite{CMS2011, ATLAS2012}.
The total measured cross section with the help of the CMS detector is  
41.1 $\pm$ 15.3 (stat) $\pm$ 5.8 (syst) $\pm$ 4.5 (lumi) pb, 
the total measured cross section with the ATLAS detector with slightly
better statistics is
54.4 $\pm$  4.0 (stat.) $\pm$  3.9 (syst.) $\pm$  2.0 (lumi.) pb. 
The more precise ATLAS result is somewhat bigger than the Standard Model
predictions of 44.4  $\pm$  2.8 pb \cite{ATLAS2012}.
The Standard Model predictions do not include several potentially
important subleading processes.

\section{Inclusive production of $W^+W^-$ pairs}
\label{sec:inclusive}

The dominant contribution of $W^+W^-$ pair production is initiated by
quark-antiquark annihilation \cite{DDS95}.
The gluon-gluon contribution to the inclusive cross section 
was calculated first in Ref.~\cite{gg_WW}.

Here we discuss the inclusive $\gamma \gamma \to W^+ W^-$
induced mechanisms. 
We calculate the contribution to the inclusive
$p p \to W^+ W^- X$ process for the first time in the literature.

\begin{figure*}
\begin{center}
\includegraphics[width=3.5cm]{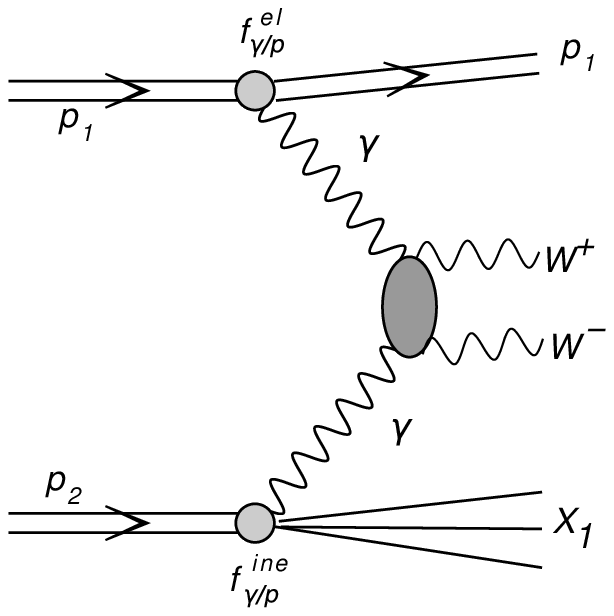}
\includegraphics[width=3.5cm]{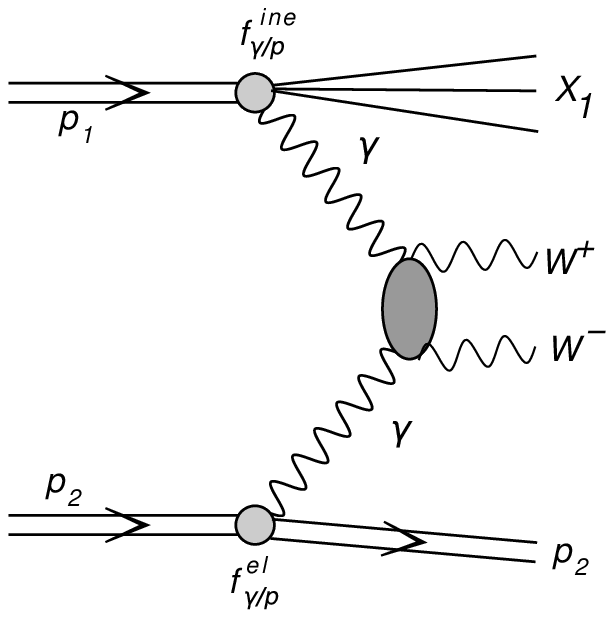}
\includegraphics[width=3.5cm]{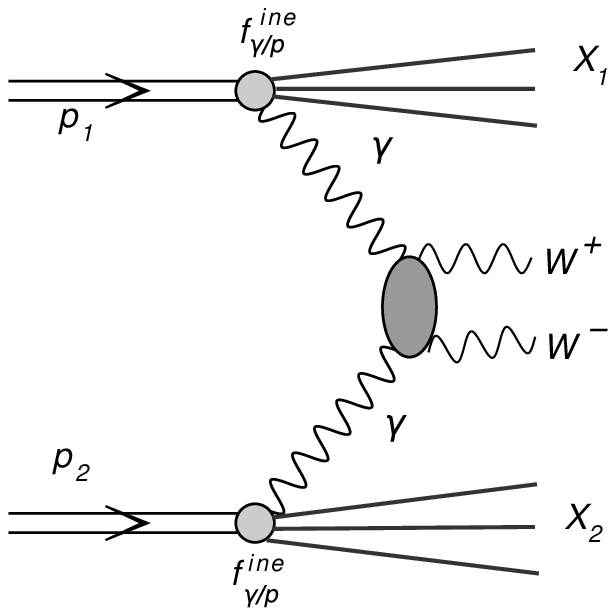}
\caption{Diagrams representing inelastic photon-photon induced mechanisms 
for the production of $W^+ W^-$ pairs.
}
\label{fig:new_diagrams}
\end{center}
\end{figure*}

If at least one photon is a constituent of the nucleon then 
the mechanisms presented in Fig.\ref{fig:new_diagrams} are possible.
In these cases at least one of the participating protons does not survive 
the $W^+ W^-$ production process. In Ref.\cite{LSR2015} we considered 
two different approaches to the problem: naive and QCD improved.

An approach how to include photons into inelastic processes 
was proposed some time ago by Martin, Roberts, 
Stirling and Thorne in Ref. \cite{MRST04}. In their approach the photon 
is treated on the same footing as quarks, antiquarks and gluons.
They proposed a QED-corrected evolution equation for the parton 
distributions of the proton \cite{MRST04}.

In leading order approximation the corresponding triple differential 
cross section for the photon-photon contribution can be written as
usually in the parton-model formalism:
\begin{eqnarray}
\frac{d \sigma^{\gamma_{1i} \gamma_{2j}}}{d y_1 d y_2 d^2p_t} 
&=& \frac{1}{16 \pi^2 {\hat s}^2}
x_1 \gamma_{1,i}(x_1,\mu^2) \; x_2 \gamma_{2j}(x_2,\mu^2) \;
\overline{|{\cal M}_{\gamma \gamma \to W^+W^-}|^2} \; .
\end{eqnarray}
where i,j = elastic,inelastic.
In the following the elastic photon fluxes are calculated using 
the Drees-Zeppenfeld parametrization \cite{DZ}, where a simple 
parametrization of the nucleon electromagnetic form factors is used.

In the case of resolved photons, the ``photonic'' quark/antiquark 
distributions in a proton must be calculated first. This can be done 
by the convolution
\begin{equation}
x f_{q/p}^{\gamma}(x) = \int_x^1 d x_{\gamma} f_{\gamma/p}(x_{\gamma},\mu_s^2) 
\left( \frac{x}{x_{\gamma}} \right) 
f \left( \frac{x}{x_{\gamma}}, \mu_h^2 \right)  \; . 
\label{convolution_resolved}       
\end{equation}

\begin{figure*}
\begin{center}
\includegraphics[width=3.5cm]{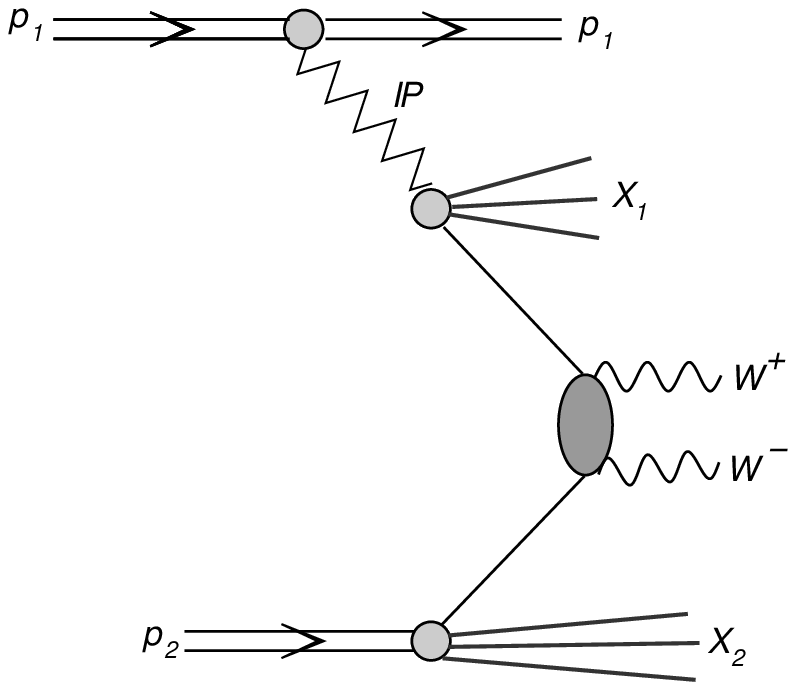}
\includegraphics[width=3.5cm]{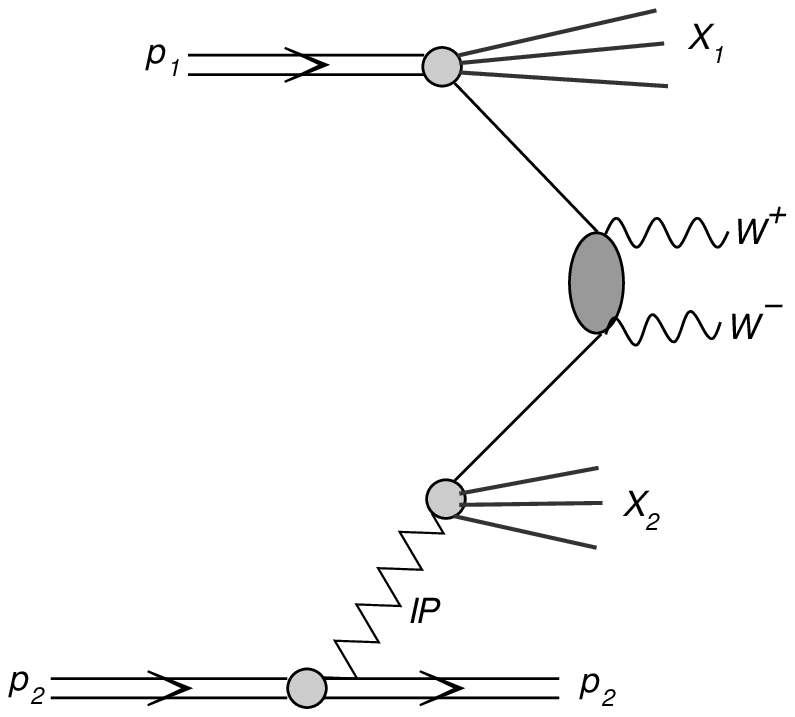}
\caption{Diagrams representing single diffractive mechanism
of the production of $W^+ W^-$ pairs.
}
\label{fig:single_diffractive}
\end{center}
\end{figure*}

Diffractive processes for $W^+ W^-$ production were not considered 
in the literature before Ref.\cite{LSR2015} 
but are potentially very important.
In the standard Ingelman-Schlein diffractive approach one assumes 
that the Pomeron has a well defined partonic structure, 
and that the hard process takes place in 
a Pomeron--proton or proton--Pomeron (single diffraction) 
or Pomeron--Pomeron (central diffraction) processes.
The mechanism of single diffractive production of $W^+ W^-$ pairs
is shown in Fig.\ref{fig:single_diffractive}.

In the present analysis we consider both pomeron and subleading reggeon
contributions. The corresponding diffractive quark distributions
are obtained by replacing the pomeron flux by the reggeon flux and
quark/antiquark distributions in the pomeron by their counterparts
in subleading reggeon(s). All other details can be found in \cite{H1}.
In the case of pomeron exchange the upper limit in the integration
over the momentum fraction carried by the pomeron/reggeon
in the convolution formula is 0.1 for pomeron and 0.2 
for reggeon exchange.
In our opinion, the Regge formalism does not apply above these limits.

Up to now we have assumed Regge factorization which is known
to be violated in hadron-hadron collisions.
It is known that these are soft interactions which lead to an extra 
production of particles which fill in the rapidity gaps related 
to pomeron exchange. The absorption effects are included here 
by multiplaying the cross section by so-called gap surrival factor.

\begin{figure*}
\begin{center}
\includegraphics[width=3.5cm]{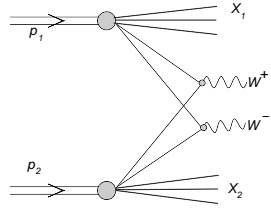}
\caption{Diagram representing double parton scattering mechanism
of the production of $W^+ W^-$ pairs.
}
\label{fig:DPS}
\end{center}
\end{figure*}

The diagram representating the double parton scattering
process is shown in Fig.\ref{fig:DPS} and the corresponding dynamics was
discussed in \cite{LSR2015}.

\section{Results}

\begin{figure}
\begin{center}
\includegraphics[width=5cm]{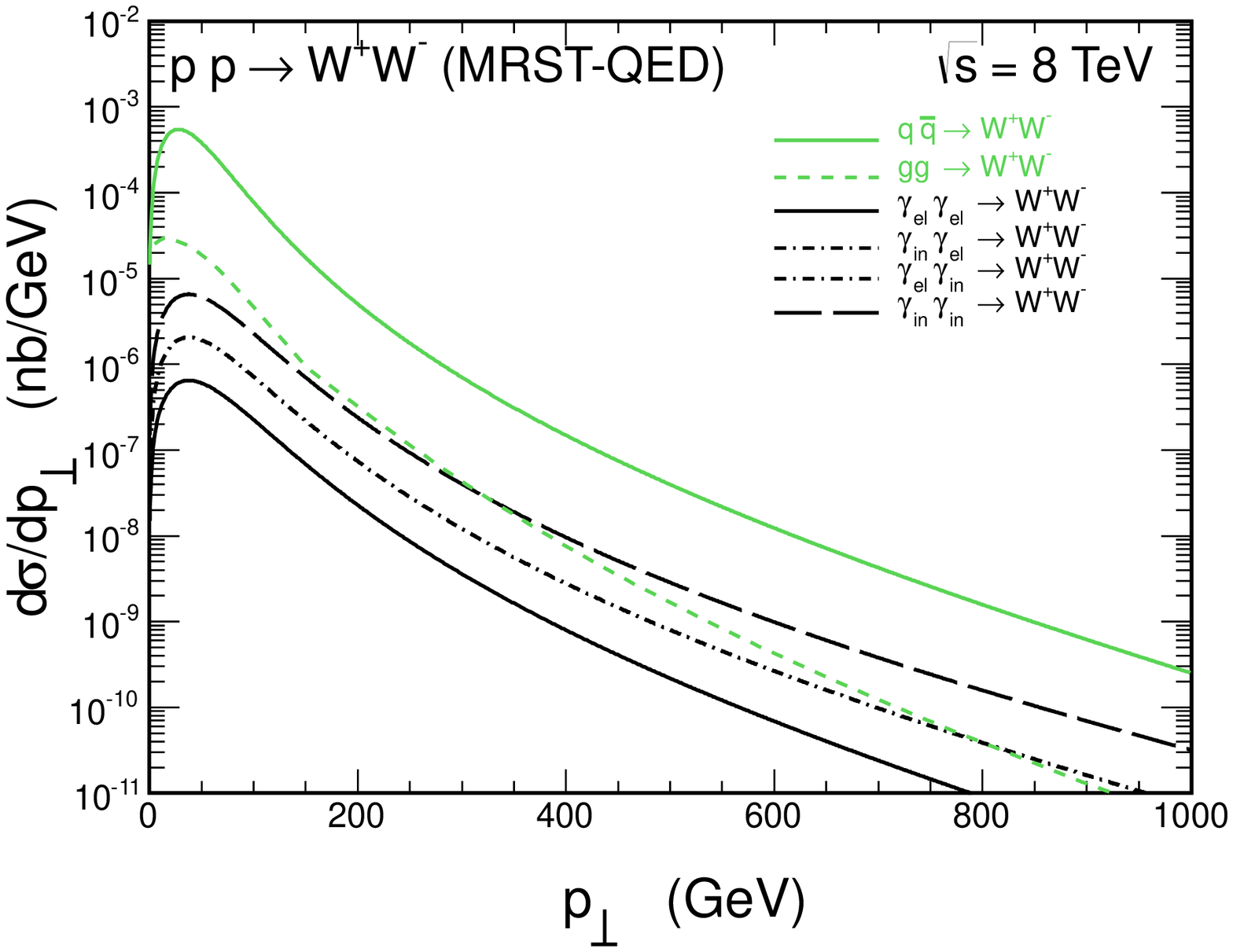}
\includegraphics[width=5cm]{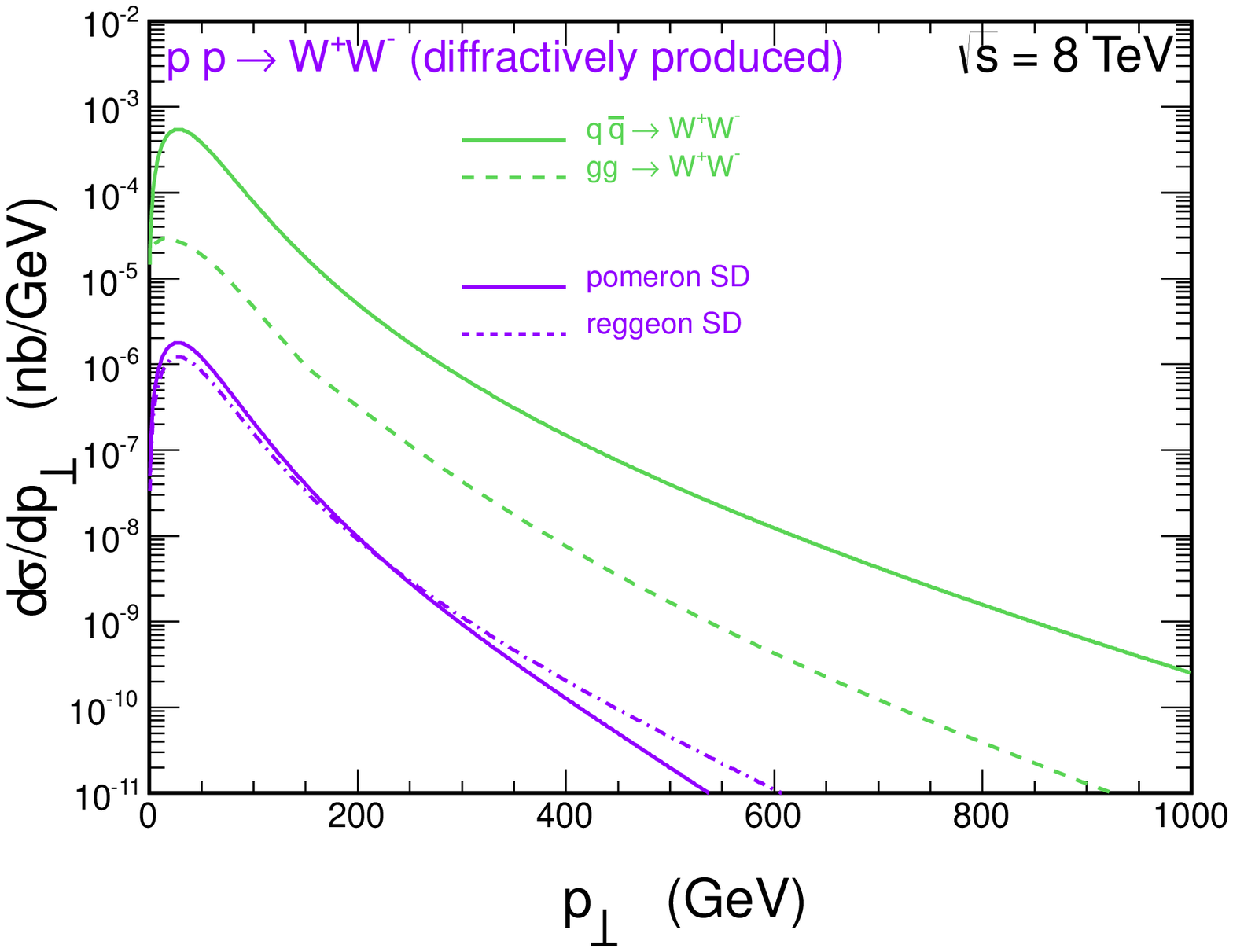}
\end{center}
\caption{ Transverse momentum distribution of $W$ bosons for 
$\sqrt{s}$ = 8 TeV.
The left panel shows all photon-photon induced processes, the
right panel the
diffractive contribution.
The diffractive cross section has been multiplied by the gap survival
factor $S_G^2 =$ 0.03.
}
\label{fig:dsig_dpt}
\end{figure}

In Fig.\ref{fig:dsig_dpt} we present distributions in the transverse momentum
of $W$ bosons. All photon-photon components have rather similar shapes.
The photon-photon contributions are somewhat harder than those for
diffractive and resolved photon mechanisms. 

%

\begin{figure}
\begin{center}
\includegraphics[width=5.5cm]{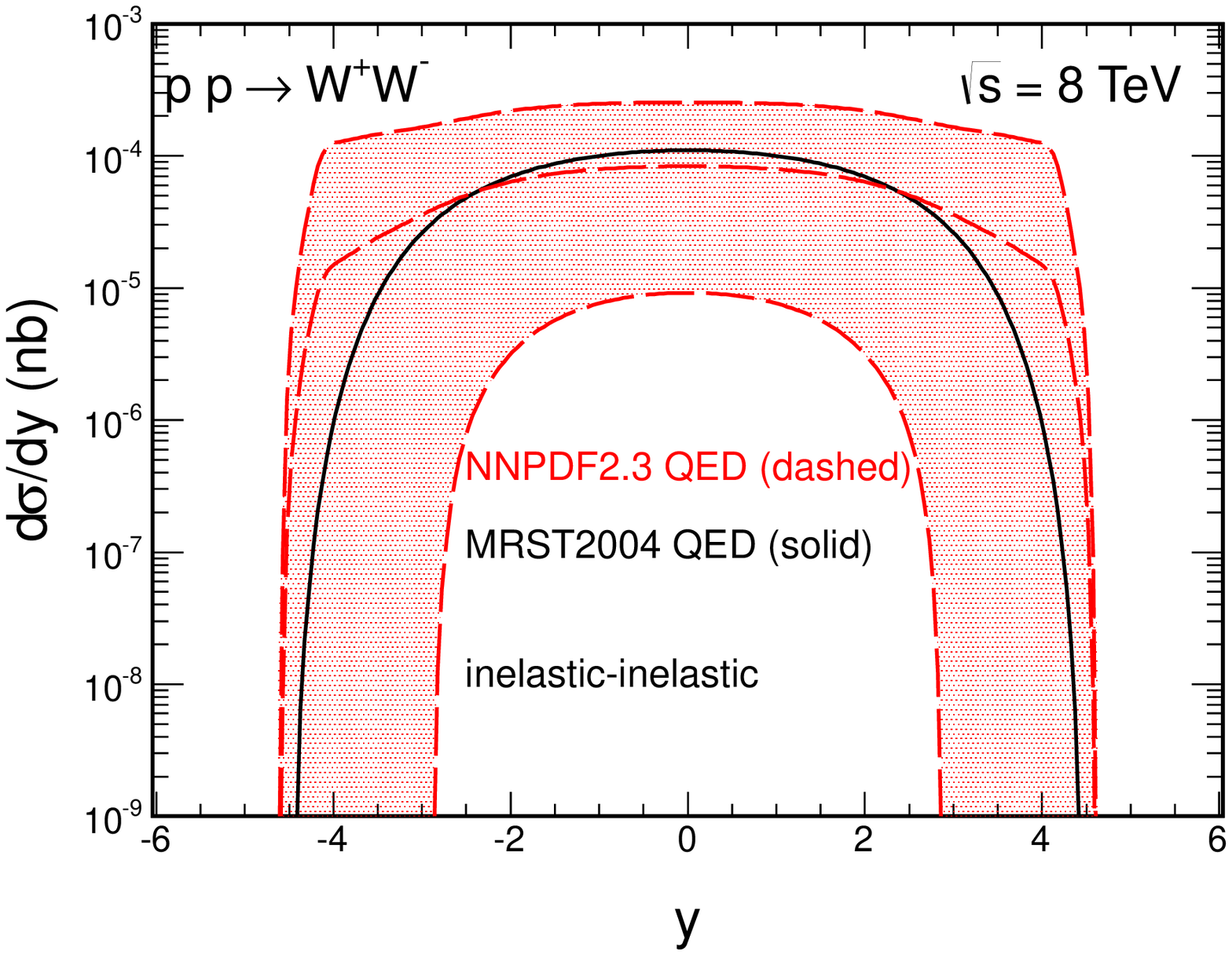}
\includegraphics[width=5.5cm]{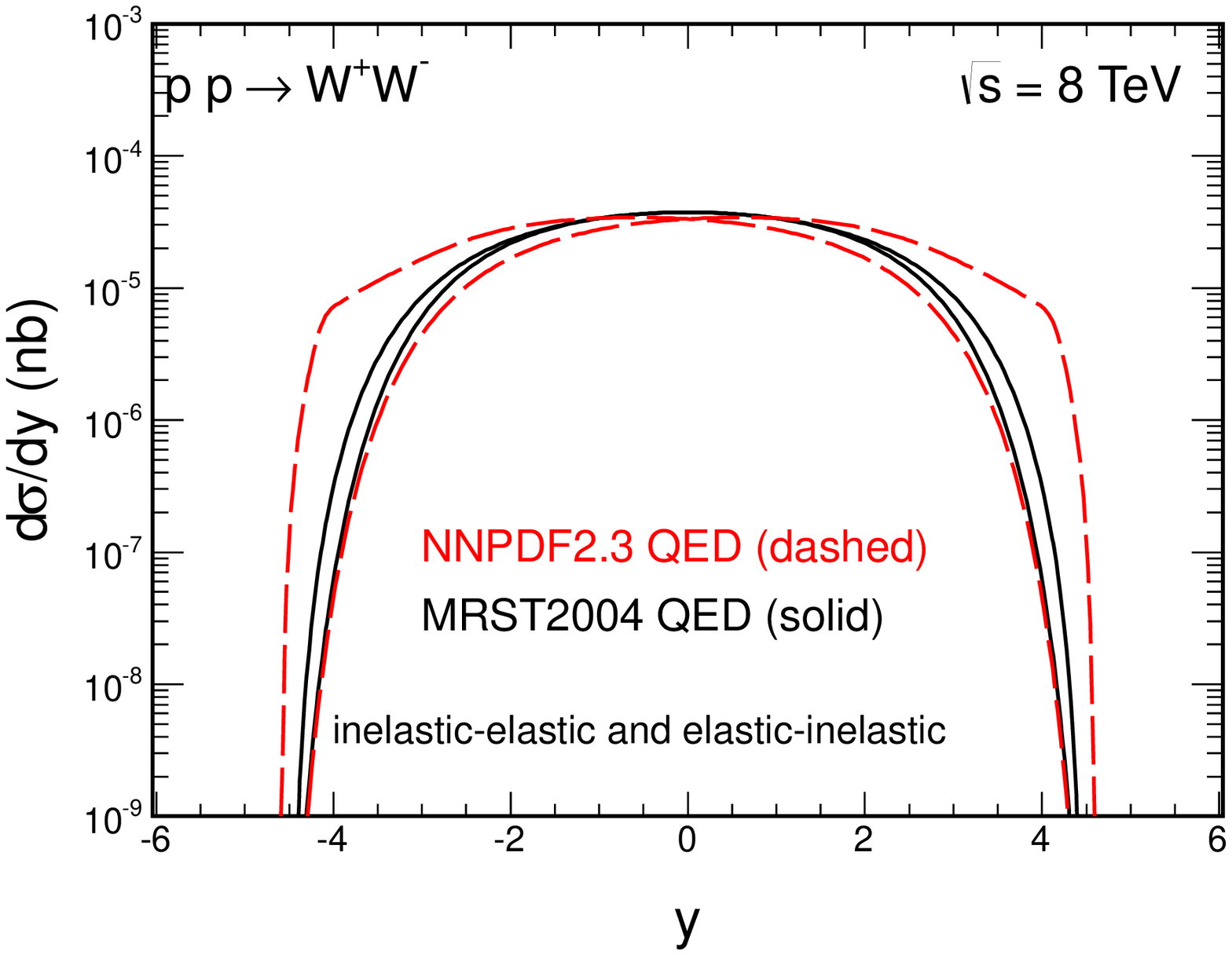}
\end{center}
\vspace{-0.7cm}
\caption {Rapidity distribution of $W$ bosons for $\sqrt{s}$ = 8 TeV
for photon-photon components with the NNPDF2.3 QED set.
In the left panel we show the dominant inelastic-inelastic component.
In addition we show uncertainty band as obtained from the NNPDF framework (one sigma).
The right panel shows elastic-inelastic and inelastic-elastic components
obtained with the NNPDF2.3 QED set. 
}
\label{fig:dsig_dy_nnpdf}
\end{figure}

We show also our predictions obtained with the NNPDF2.3 QED photon 
distributions \cite{NNPDF}, (see Fig.\ref{fig:dsig_dy_nnpdf}). 
In the left panel of Fig.\ref{fig:dsig_dy_nnpdf}
we concentrate on the biggest inelastic-inelastic contribution.
We show the statistically most probable result (middle dashed line) 
as well as one-sigma uncertainty band (shaded area). 
The uncertainty band is very large. 
This demonstrates that it is very difficult to obtain the photon
distributions from fits to experimental data. We have checked that limiting to
-2.5 $ < y_W <$ 2.5 the uncertainty band becomes relatively smaller. 
However, here we are interested mostly in the contribution in 
the whole phase space, so we leave more detailed studies
for future investigations.
The NNPDF distributions differ from those obtained with the MRST2004 QED
photon distributions at large rapidities.
In the right panel we show results for the elastic-inelastic and 
inelastic-elastic components. The most probable result obtained with 
the NNPDFs is similar as that for the MRST QED distributions. 
The elastic-inelastic and inelastic-elastic contributions 
differ one from the other more for the NNPDF than for the MRST case. 
The uncertainty bands (not shown) are also rather broad, but slightly
narrower than in the case of inelastic-inelastic component. 

\begin{figure}
\begin{center}
\includegraphics[width=5.5cm]{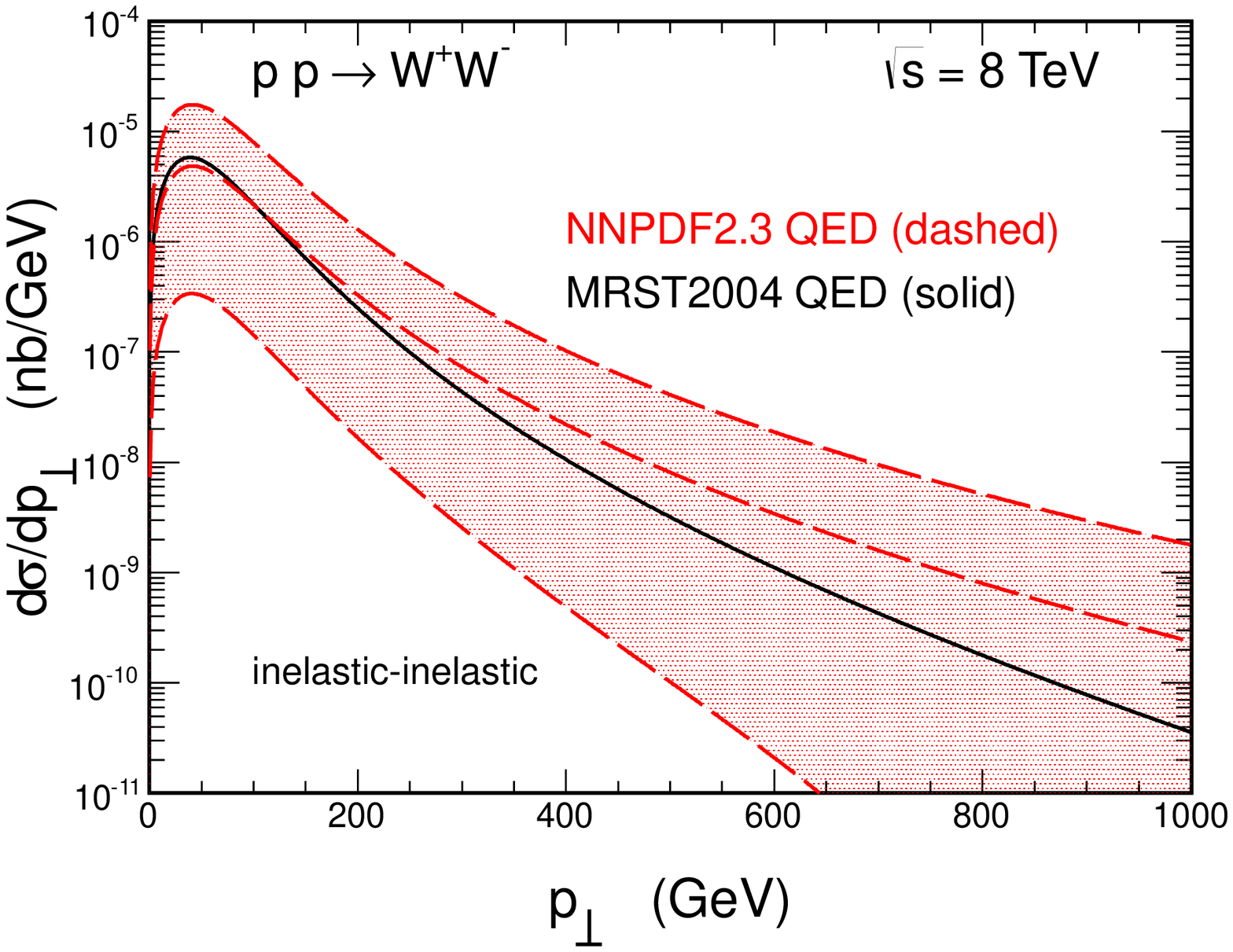}
\includegraphics[width=5.5cm]{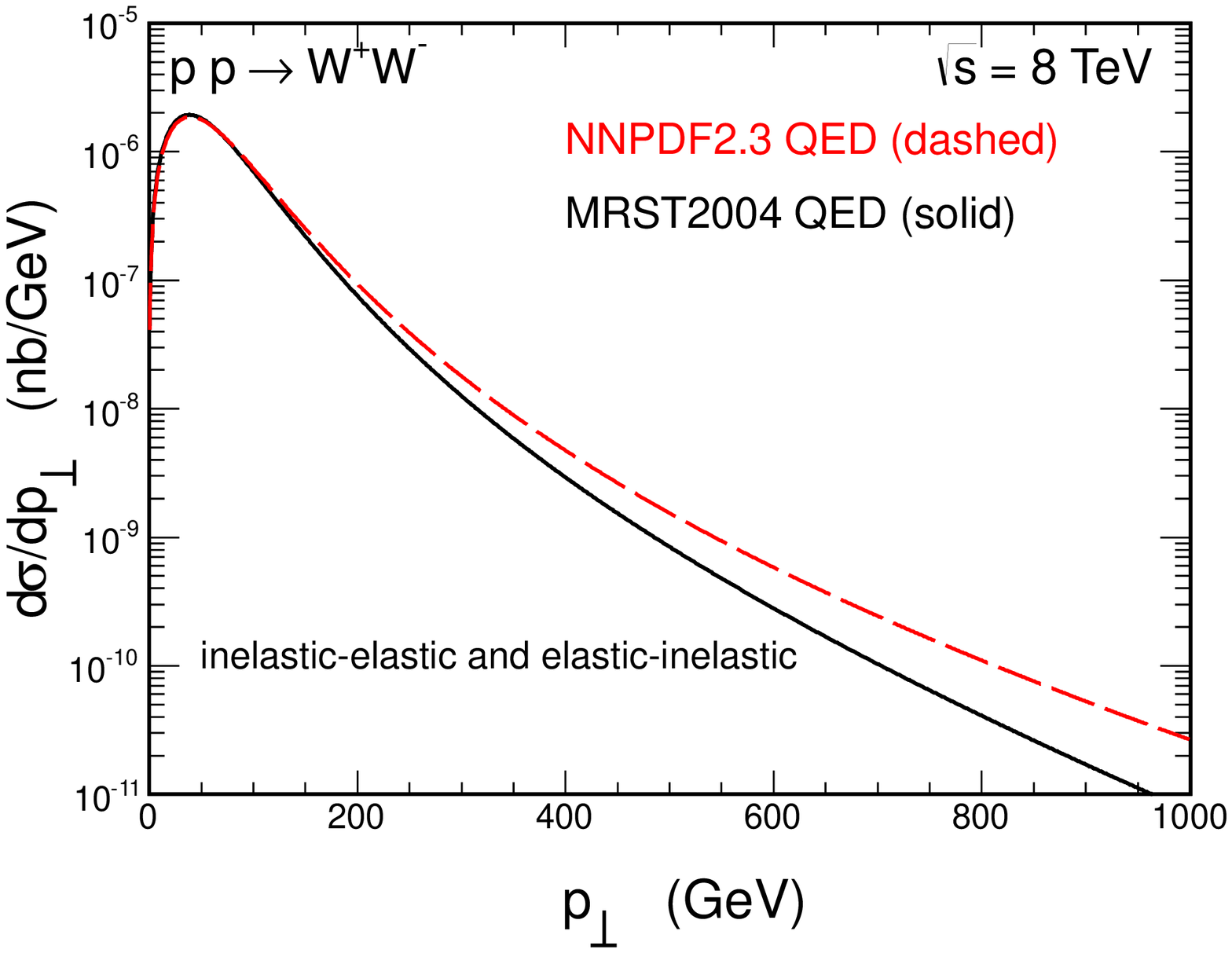}
\end{center}
\vspace{-0.7cm}
\caption {Transverse momentum distribution of $W$ bosons for $\sqrt{s}$ = 8 TeV
for photon-photon components with NNPDF2.3 QED set.
In the left panel we show the dominant inelastic-inelastic component.
In addition we show uncertainty band as obtained from the NNPDF framework (one sigma).
The right panel shows elastic-inelastic (inelastic-elastic) components
obtained with NNPDF2.3 QED set.
}
\label{fig:dsig_dpt_nnpdf}
\end{figure}

In Fig.\ref{fig:dsig_dpt_nnpdf} we show distributions
in transverse momentum of $W$ bosons for photon-photon components obtained
with the NNPDF photon distributions \cite{NNPDF}. In the left panel 
we show the dominant inelastic-inelastic component and in the right
panel elastic-inelastic (= inelastic-elastic) components. The
uncertainty band is very large, especially at large transverse momenta.

Concerning the searches for anomalous $\gamma \gamma W W$ coupling
without proton tagging as performed by the D0 and CMS collaborations, 
the ratios of the inelastic-inelastic, elastic-inelastic and 
inelastic-elastic contribution to the elastic-elastic one are crucial.

\section{Conclusions}

We have discussed nonleading, usually ignored contributions to the
production of $W^+ W^-$ pairs.

We have calculated for the first time a complete set of
photon-photon  and resolved photon-(anti)quark and (anti)quark-resolved 
photon contributions to the inclusive production of $W^+ W^-$ pairs. 
The photon-photon contributions can be classified into four topological
categories: elastic-elastic, elastic-inelastic, inelastic-elastic
and inelastic-inelastic, depending whether proton(s) survives (survive)
the emission of the photon or not. The elastic-inelastic and
inelastic-elastic contributions were calculated for the first time
in Ref.\cite{LSR2015}.
The photon-photon contributions were calculated in Ref.\cite{LSR2015} 
within the QCD-improved method using MRST(QED) and NNPDF parton distributions. 
The approach was already applied before to the production of
charged lepton pairs and $c \bar c$ pairs.
In this approach we have got
$\sigma_{ela,ela} <\sigma_{ela,ine} = \sigma_{ine,ela} < \sigma_{ine,ine}$.

We have shown in Ref.\cite{LSR2015} that including the photon 
into the evolution equation gives different results than obtained 
with more simplified approaches.
This is also a lesson for other processes known from
the literature, where photon-photon processes are possible.
This includes also some processes beyond the Standard Model mentioned in
our paper.

We have also discussed single and central diffractive contribution and
a contribution of resolved photons. Especially the single diffractive
contribution can be large. It is ignored, however, in the present analyses.
Similar situation is for double parton scattering, only mentioned here
(for details see \cite{LSR2015}).



\end{document}